\documentclass{sig-alternate}
\newfont{\mycrnotice}{ptmr8t at 7pt}
\newfont{\myconfname}{ptmri8t at 7pt}
\let\crnotice\mycrnotice%
\let\confname\myconfname%

\usepackage[german,american]{babel}
\usepackage[T1]{fontenc}
\usepackage[utf8]{inputenc}
\usepackage{times}
\usepackage{textcomp}
\usepackage{graphicx}
\usepackage{amssymb}
\usepackage{siunitx}
\usepackage{microtype} % for linebreaks in texttt

\usepackage{url}
\renewcommand{\tt}{\fontencoding{OT1}\fontfamily{cmtt}\selectfont}

\usepackage[usenames,dvipsnames]{color}
\usepackage{booktabs}
\usepackage[numbers,sectionbib]{natbib}%[round]
\makeatletter
\def\@copyrightspace{\relax}
\makeatother

\begin{document}

\title{Identifying Clickbait: A Multi-Strategy Approach Using Neural Networks\titlenote{The first four authors have equal contribution.}}
\numberofauthors{5}
\author{
\alignauthor
Vaibhav Kumar\\
\affaddr{International Institute of Information Technology Hyderabad}\\
\affaddr{vaibhav.kumar@research.iiit.ac.in}\\
\alignauthor
Dhruv Khattar\\
\affaddr{International Institute of Information Technology Hyderabad}\\
\affaddr{dhruv.khattar@research.iiit.ac.in}\\
\alignauthor
Siddhartha Gairola\\
\affaddr{International Institute of Information Technology Hyderabad}\\
\affaddr{siddhartha.gairola@research.iiit.ac.in}\\
\alignauthor
\and
Yash Kumar Lal\titlenote{The author was an intern at International Institute of Information Technology Hyderabad when this work was done.}\\
\affaddr{Manipal Institute of Technology, Manipal}\\
\affaddr{yash.kumar4@learner.manipal.edu}\\
\alignauthor
Vasudeva Varma\\
\affaddr{International Institute of Information Technology Hyderabad}\\
\affaddr{vv@iiit.ac.in}\\
}

\maketitle

\begin{abstract}
Online media outlets, in a bid to expand their reach and subsequently increase revenue through ad monetisation, have begun adopting clickbait techniques to lure readers to click on articles. The article fails to fulfill the promise made by the headline. Traditional methods for clickbait detection have relied heavily on feature engineering which, in turn, is dependent on the dataset it is built for. The application of neural networks for this task has only been explored partially. We propose a novel approach considering all information found in a social media post. We train a bidirectional LSTM with an attention mechanism to learn the extent to which a word contributes to the post's clickbait score in a differential manner. We also employ a Siamese net to capture the similarity between source and target information. Information gleaned from images has not been considered in previous approaches. We learn image embeddings from large amounts of data using Convolutional Neural Networks to add another layer of complexity to our model. Finally, we concatenate the outputs from the three separate components, serving it as input to a fully connected layer. We conduct experiments over a  test corpus of 19538 social media posts, attaining an F1 score of 65.37\% on the dataset bettering the previous state-of-the-art, as well as other proposed approaches, feature engineering or otherwise.
\end{abstract}

%%%%%%%%%%%%%%%%%%%%%%%%%%%%%%%%%%%%%%%%%%%%%%%%%%%%%%%%%%%%%%%%%%%%%%%%
\section{Introduction}

The Internet provides instant access to a wide variety of online content, news included. Formerly, users had static preferences, gravitating towards their trusted sources, incurring an unwavering sense of loyalty. The same cannot be said for current trends since users are likely to go with any source readily available to them.

In order to stay in business, news agencies have switched, in part, to a digital front. Usually, they generate revenue by (1) advertisements on their websites, or (2) a subscription based model for articles that might interest users. However, since the same information is available via multiple sources, no comment can be made on the preference of the reader. To lure in more readers and increase the number of clicks on their content, subsequently increasing their agency's revenue, writers have begun adopting a new technique - clickbait.

The concept of clickbait is formalised as something to encourage readers to click on hyperlinks based on snippets of information accompanying it, especially when those links lead to content of dubious value or interest. Clickbaiting is the intentional act of over-promising or purposely misrepresenting - in a headline, on social media, in an image, or some combination - what can be expected while reading a story on the web. It is designed to create and, consequently, capitalise on the Loewenstein information gap \cite{Loewenstein}. Sometimes, especially in cases where such headlines are found on social media, the links can redirect to a page with an unoriginal story which contains repeated or distorted facts from the original article itself.

Our engine is built on three components. The first leverages neural networks for sequential modeling of text. Article title is represented as a sequence of word vectors and each word of the title is further converted into character level embeddings. These features serve as input to a bidirectional LSTM model. An affixed attention layer allows the network to treat each word in the title in a differential manner. The next component focuses on the similarity between the article title and its actual content. For this, we generate Doc2Vec embeddings for the pair and act as input for a Siamese net, projecting them into a highly structured space whose geometry reflects complex semantic relationships. The last part of this system attempts to quantify the similarity of the attached image, if any, to the article title. Finally, the output of each component is concatenated and sent as input to a fully connected layer to generate a score for the task.

\section{Related Work}

The task of automating clickbait detection has risen to prominence fairly recently. Initial attempts for the same have worked on (1) news headlines, and (2) heavy feature engineering for the particular dataset. \cite{Biyani:2016:ASG:3015812.3015827}'s work is one of the earliest pieces of literature available in the field, focusing on an aggregation of news headlines from previously categorised clickbait and non-clickbait sources. Apart from defining different types of clickbait, they emphasise on the presence of language peculiarities exploited by writers for this purpose. These include qualitative informality metrics and use of forward references in the title to keep the reader on the hook. The first instance of detecting clickbait across social media can be traced to \cite{potthast:2016}, hand-crafting linguistic features, including a reference dictionary of clickbait phrases, over a dataset of crowdsourced tweets \cite{potthast:2017b}. However, \cite{Chakraborty2016StopCD} argued that work done specifically for Twitter had to be expanded since clickbait was available throughout the Internet, and not just social networks.

It was not until \cite{ankesh2017weused} that neural networks were tried out for the task as the authors used the same news dataset as \cite{Chakraborty2016StopCD} to develop a deep learning based model to detect clickbait. They used distributional semantics to represent article titles, and BiLSTM to model sequential data and its dependencies. Since then, \cite{PhilippeThomasCINN} has also experimented with Twitter data \cite{potthast:2017b} deploying a BiLSTM for each of the textual features (post-text, target-title, target-paragraphs, target-description, target-keywords, post-time) available in the corpus, and finally concatenating the dense output layers of the network before forwarding it to a fully connected layer. Since it was proposed in \cite{bahdanau2014neural}, the attention mechanism has been used for a variety of text-classification tasks, such as fake news detection and aspect-based sentiment analysis. \cite{YiweiZhouCDTUSN} used a self-attentive BiGRU to infer the importance of tweet tokens in predicting the annotation distribution of the task.

One common point in all the approaches yet has been the use of only textual features available in the dataset. Our model not only incorporates textual features, modeled using BiLSTM and augmented with an attention mechanism, but also considers related images for the task.

\begin{figure}[h]
   \centering
   \includegraphics[height=10cm]{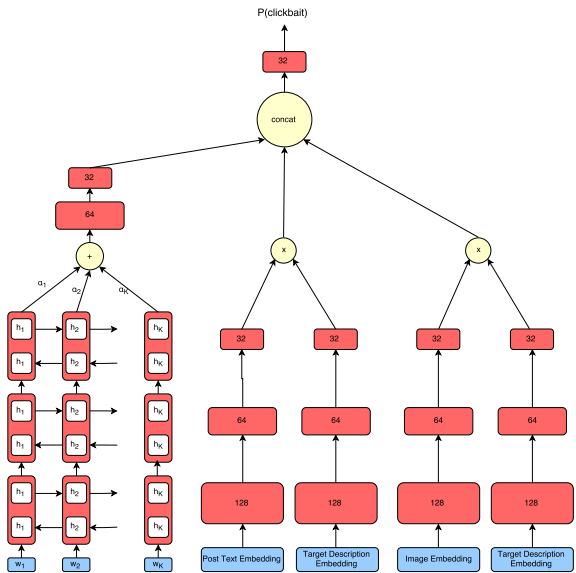}
   \caption{\label{fig:LSTMA} Model Architecture}
   \end{figure}

\section{Model Architecture}
In this section, we present our hybrid approach to clickbait detection. We first explain the three individual components followed by their fusion, which is our proposed model. These components are (1) BiLSTM with attention, (2) Siamese Network on Text Embeddings, and (3) Siamese Network on Visual Embeddings. An overview of the architecture can be seen in Figure 1.

We start with an explanation of the features used in the first component of the model.

\smallskip

\textbf{Distributed Word Embeddings}

Considering the effectiveness of distributional semantics in modeling language data, we use a pre-trained 300 dimensional Word2Vec \cite{MikolovEEWRVS} model trained over 100 billion words in the Google News corpus using the Continuous Bag of Words architecture. These map the words in a language to a high dimensional real-valued vectors to capture hidden semantic and syntactic properties of words, and are typically learned from large, unannotated text corpora. For each word in the title, we obtain its equivalent Word2Vec embeddings using the model described above.

\smallskip

\textbf{Character Level Word Embeddings}

Character level word embeddings \cite{DosSantos:2014:LCR:3044805.3045095} capture the orthographic and morphological features of a word. Apart from this, using them is a step toward mitigating the problem of out-of-vocabulary (OoV) words. In such a case, the word can be embedded by its characters using character level embedding. We follow \cite{ankesh2017weused} and first initialize a vector for every character in the corpus. The vector representation of each word is learned by applying 3 layers of a 1-dimensional Convolutional Neural Network \cite{Cun:1990:HDR:109230.109279} with ReLU non-linearity on each vector of character sequence of that word and finally max-pooling the sequence for each convolutional feature.

\smallskip

\textbf{Document Embeddings}

Doc2Vec \cite{le2014distributed} is an unsupervised approach to generate vector representations for slightly larger bodies of text, such as sentences, paragraphs and documents. It has been adapted from Word2Vec \cite{MikolovEEWRVS} which is used to generate vectors for words in large unlabeled corpora. The vectors generated by this approach come handy in tasks like calculating similarity metrics for sentences, paragraphs and documents. In sequential models like RNNs, the word sequence is captured in the generated sentence vectors. However, in Doc2Vec, the representations are order independent. We use GenSim \cite{radim} to learn 300 dimensional Doc2Vec embeddings for each target description and post title available.

\smallskip

\textbf{Pre-trained CNN Features}

As seen in various visual understanding problems recently, image descriptors trained using Convolutional Neural Networks over large amounts of data such as ImageNet have proven to be very effective. The implicit learning of spatial layout and object semantics in the later layers of the network from very large datasets has contributed to the success of these features. We use a pre-trained network of VGG-19 architecture \cite{VeryDeepCNNImageRec} trained over the ImageNet database (ILSVRC-2012) and extract CNN features. We use the output of the fully-connected layer (FC7), which has 4096 dimensions, as feature representations for our architecture.

\smallskip

We now go into detail about the components of the model, individual and combined, and how the parameters are learned.

\smallskip

\subsection{Bidirectional LSTM with Attention}

Recurrent Neural Network (RNN) is a class of artificial neural networks which utilizes sequential information and maintains history through its intermediate layers. A standard RNN has an internal state whose output at every time-step which can be expressed in terms of that of previous time-steps. However, it has been seen that standard RNNs suffer from a problem of vanishing gradients \cite{hochreiter1997long}. This means it will not be able to efficiently model dependencies and interactions between words that are a few steps apart. LSTMs are able to tackle this issue by their use of gating mechanisms. For each record in the dataset, the content of the post as well as the content of the related web page is available. We convert the words from the title of both attributes into the previously mentioned types of embeddings to act as input to our bidirectional LSTMs.

$(\overrightarrow{h}_1, \overrightarrow{h}_2, \dots, \overrightarrow{h}_R)$ represent forward states of the LSTM and its state updates satisfy the following equations:
\begin{equation}
    \big[\overrightarrow{f_t},\overrightarrow{i_t},\overrightarrow{o_t}\big] = \sigma \big[ \overrightarrow{W} \big[\overrightarrow{h}_{t-1},\overrightarrow{r_t}\big] + \overrightarrow{b}\big]
\end{equation}
\begin{equation}
    \overrightarrow{l_t} = \tanh \big[\overrightarrow{V} \big[\overrightarrow{h}_{t-1}, \overrightarrow{r_t}\big] + \overrightarrow{d}\big]
\end{equation}
\begin{equation}
    \overrightarrow{c_t} = \overrightarrow{f_t} \cdot \overrightarrow{c}_{t-1} + \overrightarrow{i_t} \cdot \overrightarrow{l_t}
\end{equation}
\begin{equation}
    \overrightarrow{h_t} = \overrightarrow{o_t} \cdot \tanh(\overrightarrow{c_t})
\end{equation}
here $\sigma$ is the logistic sigmoid function, $\overrightarrow{f_t}$, $\overrightarrow{i_t}$, $\overrightarrow{o_t}$ represent the forget, input and output gates respectively. $\overrightarrow{r_t}$ denotes the input at time $t$ and $\overrightarrow{h_t}$ denotes the latent state, $\overrightarrow{b_t}$ and $\overrightarrow{d_t}$ represent the bias terms. The forget, input and output gates control the flow of information throughout the sequence. $\overrightarrow{W}$ and $\overrightarrow{V}$ are matrices which represent the weights associated with the connections.

 $(\overleftarrow{h}_1, \overleftarrow{h}_2, \dots, \overleftarrow{h}_R)$ denote the backward states and its updates can be computed similarly.
 
 % add backward state update equations if needed
 
 The number of bidirectional LSTM units is set to a constant \textit{K}, which is the maximum length of all title lengths of records used in training. The forward and backward states are then concatenated to obtain $(h_1, h_2, \dots, h_K)$, where
\begin{equation}
    h_i = \begin{bmatrix}
               \overrightarrow{h}_i \\
               \overleftarrow{h}_i
          \end{bmatrix}
\end{equation}
Finally, we are left with the task of figuring out the significance of each word in the sequence i.e. how much a particular word influences the clickbait-y nature of the post. The effectiveness of attention mechanisms have been proven for the task of neural machine translation \cite{bahdanau2014neural} and it has the same effect in this case. The goal of attention mechanisms in such tasks is to derive context vectors which capture relevant source side information and help predict the current target word. The sequence of annotations generated by the encoder to come up with a context vector capturing how each word contributes to the record's clickbait quotient is of paramount importance to this model. In a typical RNN encoder-decoder framework \cite{bahdanau2014neural}, a context vector is generated at each time-step to predict the target word. However, we only need it for calculation of context vector for a single time-step.
\begin{equation}
c_{attention} = \sum_{j=1}^{K}\alpha_jh_j
\end{equation}
where, $h_1$,\dots,$h_K$ represents the sequence of annotations to which the encoder maps the post title vector and each $\alpha_j$ represents the respective weight corresponding to each annotation $h_j$. This component is represented on the leftmost in Figure 1.

\subsection{Siamese Net with Text Embeddings}

The second component of our model is a Siamese net \cite{siamese} over two textual features in the dataset. Siamese networks are designed around having symmetry and it is important because it's required for learning a distance metric. We use them to find the similarity between the title of the record and its target description. The words in the title and in the target description are converted into their respective Doc2Vec embeddings and concatenated, after which they are considered as input into a Siamese network. A visual representation of this can be found in the middle of Figure 1.

\subsection{Siamese Neural Network with Visual Embeddings}

The final component of our hybrid model is also a Siamese net. However, it considers visual information available in the dataset, and sets our model apart from other approaches in this field. The relevance of the image attached to the post can be quantified by capturing its similarity with the target description. The VGG-19 architecture outputs a 4096 dimensional vector for each image which, in turn, is fed as input into a dense layer to convert each representation to a 300 dimensional vector. This serves as one input to the visual Siamese net. The target description is converted into its 300 dimensional vector representation by passing it through the pre-trained Doc2Vec model, which acts as the second input for the network. It is the rightmost part of Figure 1.

\subsection{Fusion of the components}

To combine the components and complete our hybrid model, the output from each of the three parts is concatenated and subsequently acts as input for a fully connected layer. This layer finally gives as its output the probability/extent that a post, together with its related information, can be considered clickbait.

\subsection{Learning the Parameters}

We use binary cross-entropy as the loss optimization function for our model. The cross-entropy method \cite{deBoer2005} is an iterative procedure where each iteration can be divided into two stages:

(1) Generate a random data sample (vectors, trajectories etc.) according to a specified mechanism.

(2) Update the parameters of the random mechanism based on the data to produce a "better" sample in the next iteration.

\section{Evaluation Results}

The model was evaluated over a collection of 19538 social media posts \cite{potthast:2017b}, each containing supplementary information like target description, target keywords and linked images. We performed our experiments with the aim of increasing the accuracy and F1 score of the model. Other metrics like mean squared error (MSE) were also considered.

\subsection{Training}

We randomly partition the training set into training and validation set in a 4:1 ratio. This ensures that the two sets do not overlap. The model hyperparameters are tuned over the validation set. We initialise the fully connected network weights with the uniform distribution in the range $-\sqrt{{6}/{(fanin + fanout)}}$ and $\sqrt{{6}/{(fanin + fanout)}}$ \cite{glorot2010understanding}. We used a batch size of 256 and adadelta \cite{zeiler2012adadelta} as a gradient based optimizer for learning the parameters of the model.

\subsection{Comparison with other models}

In Table 1, we compare our model with the existing state-of-the-art for the dataset used and other models which have employed similar techniques to accomplish the task. Calculation and comparison across these metrics was conducted on TIRA \cite{potthast:2016}, a platform that offers evaluation as a service. It is clear that our proposed model outperforms the previous feature engineering benchmark and other work done in the field both in terms of F1 score and accuracy of detection.

\begin{table}
\caption{Model Performance Comparison}
\begin{tabular}{|c|c|c|} \hline
\textbf{Model}&\textbf{F1 Score}&\textbf{Accuracy}\\ \hline
Proposed Hybrid Approach & \textbf{0.65}& \textbf{83.53}\%\\ \hline
BiLSTM \cite{ankesh2017weused} & 0.61& 83.28\%\\ \hline
Feature Engineering Baseline \cite{potthast:2016} & 0.55& 83.24\%\\ \hline
Concatenated NN Architecture \cite{PhilippeThomasCINN} & 0.39& 74\%\\
\hline\end{tabular}
\end{table}

\section{Conclusion}

In this work, we have come up with a multi-strategy approach to tackle the problem of clickbait detection across the Internet. Our model takes into account both textual and image features, a multimedia approach, to score the classify headlines. A neural attention mechanism is utilised over \cite{ankesh2017weused} to improve its performance, simultaneously adding Siamese nets for scoring similarity between different attributes of the post. To build on this approach, we would like to explore better image embedding techniques to better relate it to the article.

%%%%%%%%%%%%%%%%%%%%%%%%%%%%%%%%%%%%%%%%%%%%%%%%%%%%%%%%%%%%%%%%%%%%%%%%
\begin{raggedright}
\bibliography{clickbait17-notebook-lit}

\begin{thebibliography}{20}
\providecommand{\natexlab}[1]{#1}
\providecommand{\url}[1]{\texttt{#1}}
\expandafter\ifx\csname urlstyle\endcsname\relax
  \providecommand{\doi}[1]{doi: #1}\else
  \providecommand{\doi}{doi: \begingroup \urlstyle{rm}\Url}\fi

\bibitem[Anand et~al.(2017)Anand, Chakraborty, and Park]{ankesh2017weused}
A.~Anand, T.~Chakraborty, and N.~Park.
\newblock {We used Neural Networks to Detect Clickbaits: You won't believe what
  happened Next!}
\newblock In \emph{Advances in Information Retrieval. 39th European Conference
  on IR Research (ECIR 17)}, Lecture Notes in Computer Science. Springer, 2017.

\bibitem[Bahdanau et~al.(2014)Bahdanau, Cho, and Bengio]{bahdanau2014neural}
D.~Bahdanau, K.~Cho, and Y.~Bengio.
\newblock Neural machine translation by jointly learning to align and
  translate.
\newblock \emph{arXiv preprint arXiv:1409.0473}, 2014.

\bibitem[Biyani et~al.(2016)Biyani, Tsioutsiouliklis, and
  Blackmer]{Biyani:2016:ASG:3015812.3015827}
P.~Biyani, K.~Tsioutsiouliklis, and J.~Blackmer.
\newblock "8 amazing secrets for getting more clicks": Detecting clickbaits in
  news streams using article informality.
\newblock In \emph{Proceedings of the Thirtieth AAAI Conference on Artificial
  Intelligence}, AAAI'16, pages 94--100. AAAI Press, 2016.
\newblock URL \url{http://dl.acm.org/citation.cfm?id=3015812.3015827}.

\bibitem[Chakraborty et~al.(2016)Chakraborty, Paranjape, Kakarla, and
  Ganguly]{Chakraborty2016StopCD}
A.~Chakraborty, B.~Paranjape, S.~Kakarla, and N.~Ganguly.
\newblock Stop clickbait: Detecting and preventing clickbaits in online news
  media.
\newblock \emph{2016 IEEE/ACM International Conference on Advances in Social
  Networks Analysis and Mining (ASONAM)}, pages 9--16, 2016.

\bibitem[Cun et~al.(1990)Cun, Boser, Denker, Howard, Habbard, Jackel, and
  Henderson]{Cun:1990:HDR:109230.109279}
Y.~L. Cun, B.~Boser, J.~S. Denker, R.~E. Howard, W.~Habbard, L.~D. Jackel, and
  D.~Henderson.
\newblock Advances in neural information processing systems 2.
\newblock chapter Handwritten Digit Recognition with a Back-propagation
  Network, pages 396--404. Morgan Kaufmann Publishers Inc., San Francisco, CA,
  USA, 1990.
\newblock ISBN 1-55860-100-7.
\newblock URL \url{http://dl.acm.org/citation.cfm?id=109230.109279}.

\bibitem[de~Boer et~al.(2005)de~Boer, Kroese, Mannor, and
  Rubinstein]{deBoer2005}
P.-T. de~Boer, D.~P. Kroese, S.~Mannor, and R.~Y. Rubinstein.
\newblock A tutorial on the cross-entropy method.
\newblock \emph{Annals of Operations Research}, 134\penalty0 (1):\penalty0
  19--67, Feb 2005.
\newblock ISSN 1572-9338.
\newblock \doi{10.1007/s10479-005-5724-z}.
\newblock URL \url{https://doi.org/10.1007/s10479-005-5724-z}.

\bibitem[Dos~Santos and Zadrozny(2014)]{DosSantos:2014:LCR:3044805.3045095}
C.~N. Dos~Santos and B.~Zadrozny.
\newblock Learning character-level representations for part-of-speech tagging.
\newblock In \emph{Proceedings of the 31st International Conference on
  International Conference on Machine Learning - Volume 32}, ICML'14, pages
  II--1818--II--1826. JMLR.org, 2014.
\newblock URL \url{http://dl.acm.org/citation.cfm?id=3044805.3045095}.

\bibitem[Glorot and Bengio(2010)]{glorot2010understanding}
X.~Glorot and Y.~Bengio.
\newblock Understanding the difficulty of training deep feedforward neural
  networks.
\newblock In \emph{Aistats}, volume~9, pages 249--256, 2010.

\bibitem[Hochreiter and Schmidhuber(1997)]{hochreiter1997long}
S.~Hochreiter and J.~Schmidhuber.
\newblock Long short-term memory.
\newblock \emph{Neural computation}, 9\penalty0 (8):\penalty0 1735--1780, 1997.

\bibitem[Le and Mikolov(2014)]{le2014distributed}
Q.~Le and T.~Mikolov.
\newblock Distributed representations of sentences and documents.
\newblock In \emph{Proceedings of the 31st International Conference on Machine
  Learning (ICML-14)}, pages 1188--1196, 2014.

\bibitem[Loewenstein(1994)]{Loewenstein}
G.~Loewenstein.
\newblock The psychology of curiosity: A review and reinterpretation.
\newblock 116:\penalty0 75--98, 07 1994.

\bibitem[Mikolov et~al.(2013)Mikolov, Chen, Corrado, and Dean]{MikolovEEWRVS}
T.~Mikolov, K.~Chen, G.~Corrado, and J.~Dean.
\newblock Efficient estimation of word representations in vector space.
\newblock \emph{CoRR}, abs/1301.3781, 2013.
\newblock URL \url{http://arxiv.org/abs/1301.3781}.

\bibitem[Neculoiu et~al.(2016)Neculoiu, Versteegh, and Rotaru]{siamese}
P.~Neculoiu, M.~Versteegh, and M.~Rotaru.
\newblock Learning text similarity with siamese recurrent networks.
\newblock 01 2016.

\bibitem[Potthast et~al.(2016)Potthast, K{\"o}psel, Stein, and
  Hagen]{potthast:2016}
M.~Potthast, S.~K{\"o}psel, B.~Stein, and M.~Hagen.
\newblock {Clickbait Detection}.
\newblock In N.~Ferro, F.~Crestani, M.-F. Moens, J.~Mothe, F.~Silvestri, G.~{Di
  Nunzio}, C.~Hauff, and G.~Silvello, editors, \emph{Advances in Information
  Retrieval. 38th European Conference on IR Research (ECIR 16)}, volume 9626 of
  \emph{Lecture Notes in Computer Science}, pages 810--817, Berlin Heidelberg
  New York, Mar. 2016. Springer.

\bibitem[Potthast et~al.(2017)Potthast, Gollub, Komlossy, Schuster, Wiegmann,
  Garces, Hagen, and Stein]{potthast:2017b}
M.~Potthast, T.~Gollub, K.~Komlossy, S.~Schuster, M.~Wiegmann, E.~Garces,
  M.~Hagen, and B.~Stein.
\newblock {Crowdsourcing a Large Corpus of Clickbait on Twitter}.
\newblock In \emph{{(to appear)}}, 2017.

\bibitem[{\v R}eh{\r u}{\v r}ek and Sojka(2010)]{radim}
R.~{\v R}eh{\r u}{\v r}ek and P.~Sojka.
\newblock {Software Framework for Topic Modelling with Large Corpora}.
\newblock In \emph{{Proceedings of the LREC 2010 Workshop on New Challenges for
  NLP Frameworks}}, pages 45--50, Valletta, Malta, May 2010. ELRA.
\newblock \url{http://is.muni.cz/publication/884893/en}.

\bibitem[Simonyan and Zisserman(2014)]{VeryDeepCNNImageRec}
K.~Simonyan and A.~Zisserman.
\newblock Very deep convolutional networks for large-scale image recognition.
\newblock \emph{CoRR}, abs/1409.1556, 2014.

\bibitem[Thomas(2017)]{PhilippeThomasCINN}
P.~Thomas.
\newblock Clickbait identification using neural networks.
\newblock \emph{CoRR}, abs/1710.08721, 2017.
\newblock URL \url{http://arxiv.org/abs/1710.08721}.

\bibitem[Zeiler(2012)]{zeiler2012adadelta}
M.~D. Zeiler.
\newblock Adadelta: an adaptive learning rate method.
\newblock \emph{arXiv preprint arXiv:1212.5701}, 2012.

\bibitem[Zhou(2017)]{YiweiZhouCDTUSN}
Y.~Zhou.
\newblock Clickbait detection in tweets using self-attentive network.
\newblock \emph{CoRR}, abs/1710.05364, 2017.
\newblock URL \url{http://arxiv.org/abs/1710.05364}.

\end{thebibliography}
\end{raggedright}
\end{document}